\documentclass[11pt,twoside]{article}
\usepackage{asp2006}
\usepackage{graphicx}
\usepackage{txfonts}
\begin{document}

\title{Extended sources from deep GMRT 150 MHz observations}

\author{S.~J.~George$^1$ and C.~H.~Ishwara-Chandra$^2$}

\affil{$^1$School of Physics \& Astronomy, University of Birmingham, UK\\
       $^2$National Centre for Radio Astrophysics, TIFR, Ganeshkhind, Pune
           411 007, India}

\begin{abstract}
We present results of deep 150 MHz observations with the GMRT which show
several extended radio sources with a range of morphologies.  These sources
have then further followed up at higher frequencies (610 and 1400~MHz) with the
GMRT. GMRT J0137+4121 was a candidate double--double radio galaxy for which we
have also used the VLA-A array at C band to resolve the core. These
observations have allowed us to determine that this source is a normal radio
galaxy with a core and a one sided jet. Prominent amongst the other extended
sources is the giant radio galaxy, 4C39.04.
\end{abstract}

\section{Introduction}

The field of Upsilon Andromeda was observed with the GMRT at 150 MHz as part of
a search for extrasolar planets (Winterhalter et al.\ 2005). The rms noise of
the images was $\sim 1.5$ mJy beam$^{-1}$ with a resolution of $\sim 20''$.
Though the extrasolar planet was not detected, the low rms noise along with the
superior resolution at the low frequency of 150 MHz enabled us to detect a few
hundred-background radio sources. Cross-correlation of these sources at higher
frequencies was used to obtain ultra-steep spectrum (USS) sources, which are
potential candidates for high-redshift radio galaxies (Ishwara-Chandra \&
Marathe 2007). Further investigation of the image revealed several radio
sources with a range of morphologies like compact double, diffuse, one-sided,
etc. Here we present the images at 150 MHz for some of the sources and results
of follow-up observations of one interesting radio source which appeared like a
double--double radio galaxy.

\section{Observations and data analysis}

The 150 MHz observations of this field were carried with GMRT on July 27, 2004
using a bandwidth of 5.5 MHz. GMRT consists of 30 antennas, each of 45 meter
diameter located 90 km from Pune, India and operates at five frequency bands
from 150 MHz to 1450 MHz. For determining the flux scale, the standard flux
calibrator 3C48 was observed and the flux at 150 MHz was estimated using Baars
formula. The data of Upsilon Andromeda was analysed in {\tt AIPS} using
standard procedures for wide field imaging with appropriate care taken while
averaging channels to avoid bandwidth smearing. A few iterations of phase-only
self calibration and one iteration of amplitude and phase self-calibration was
applied to improve the image quality. The final rms noise achieved at 150 MHz
near the centre of the field was $\sim 1.5$ mJy~beam$^{-1}$ and resolution of
$\sim 20''$.

\begin{figure}
\centerline{\includegraphics[height=5.5cm]{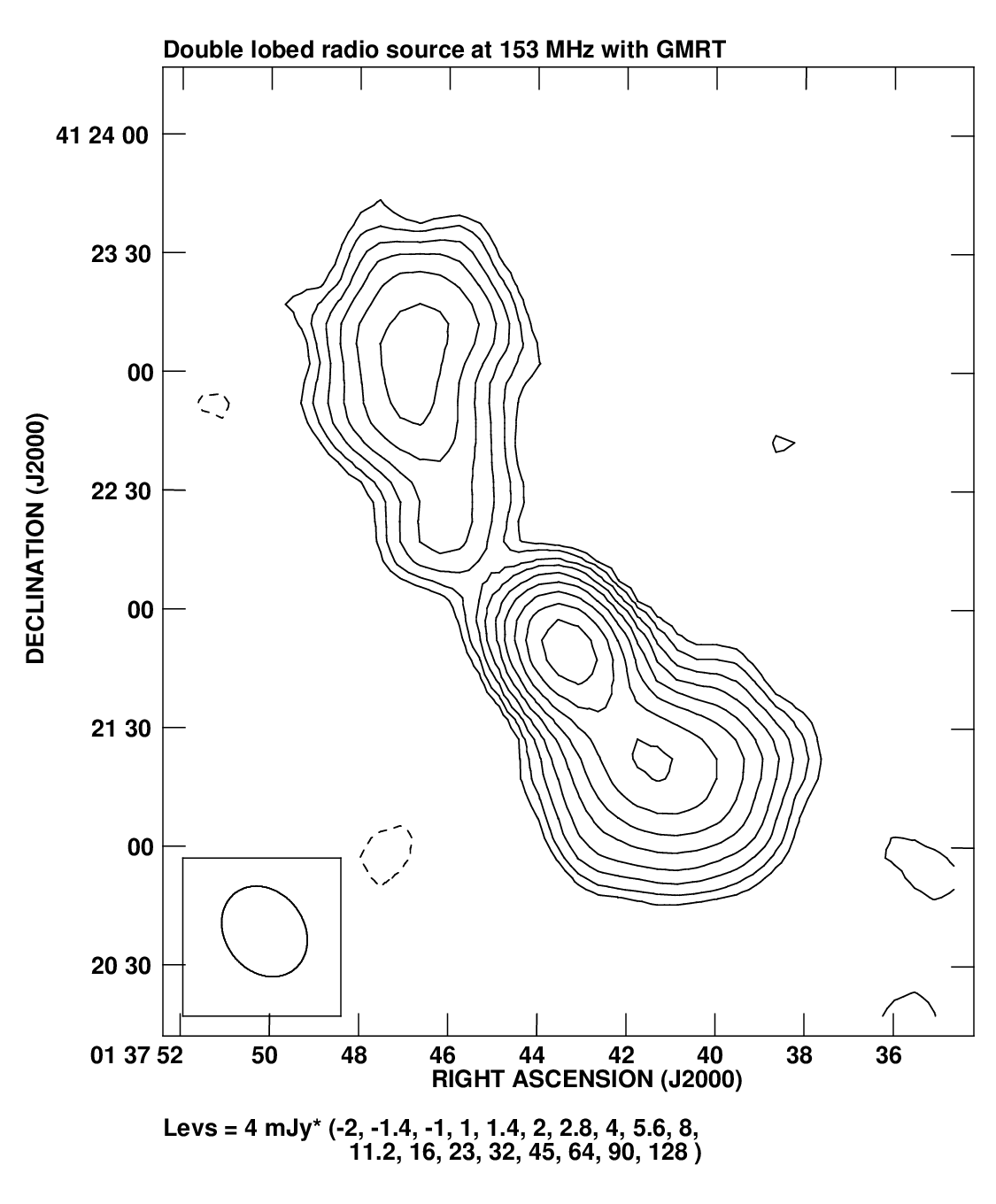}
   \kern5pt \includegraphics[height=5.5cm]{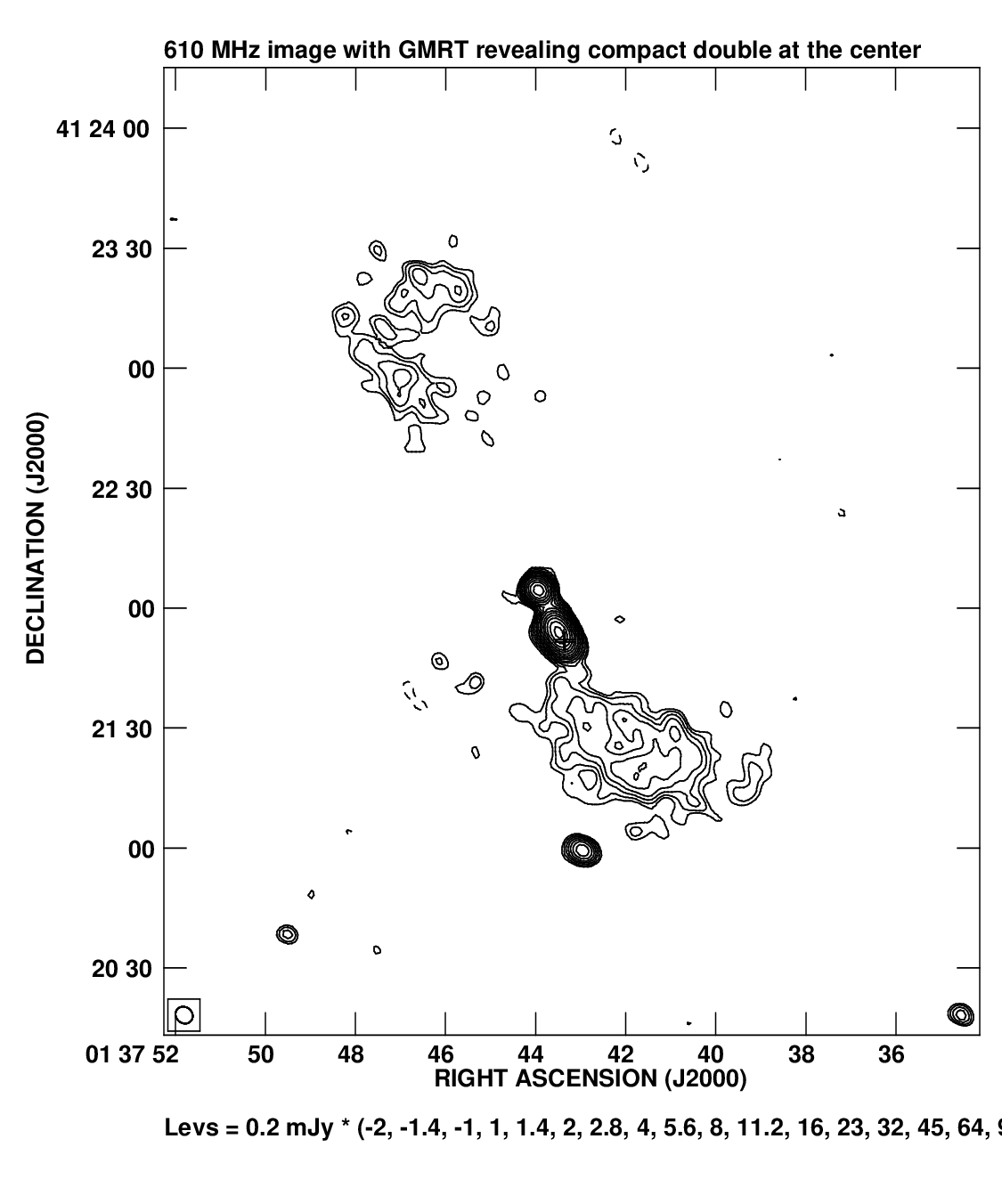}
  \kern-5pt \includegraphics[height=5.5cm]{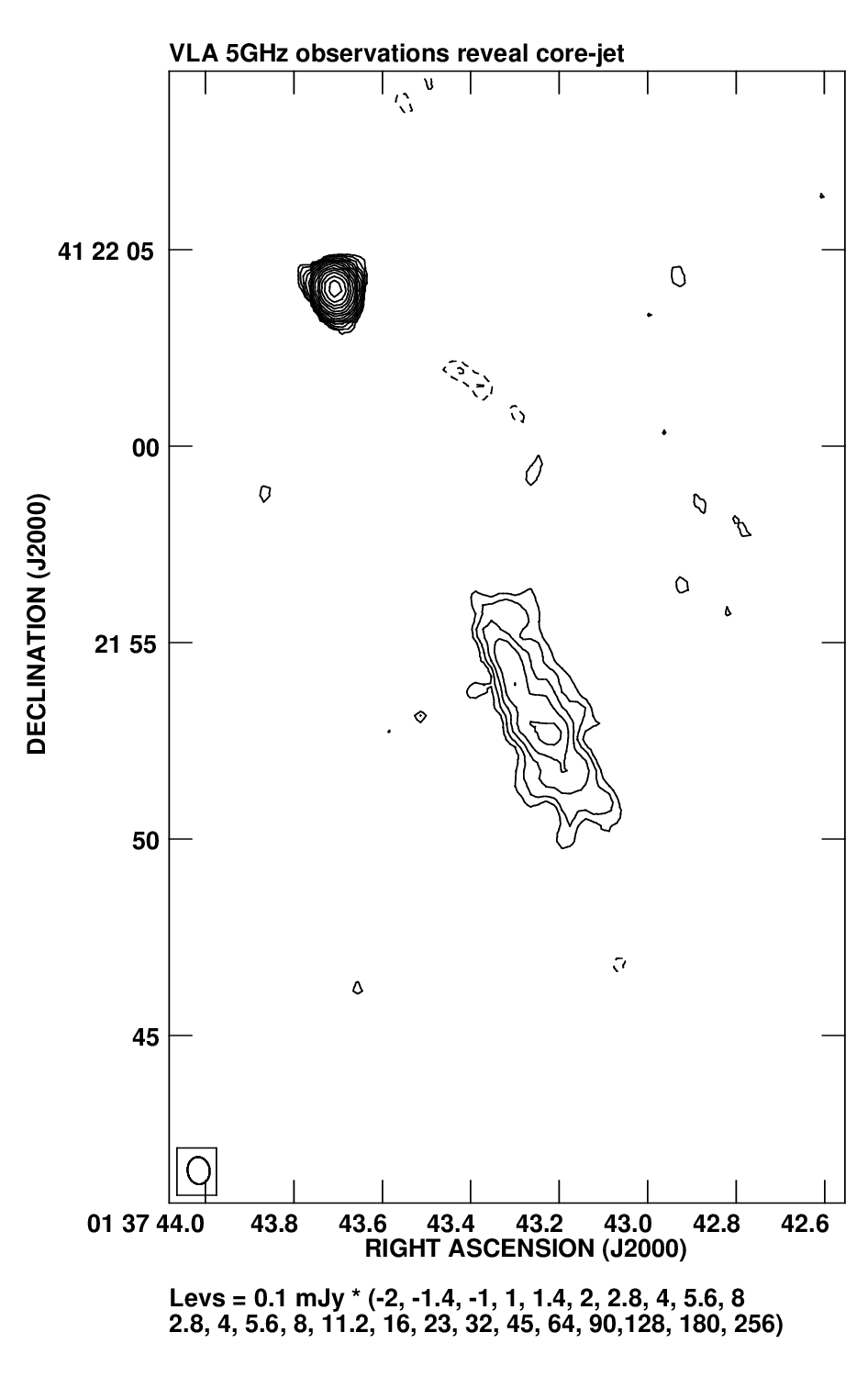}}
\caption{GMRT J0137+4121. {\it Left:} GMRT 150 MHz image, indicating partially
resolved source at the centre. {\it Middle:} At 610 MHz, the image indeed
showed possibility of a double; {\it Right:} The VLA 5 GHz observation resolves
the compact double into core and one sided detached jet.}
\end{figure}

\begin{figure}
\centerline{\includegraphics[angle=270,width=8cm]{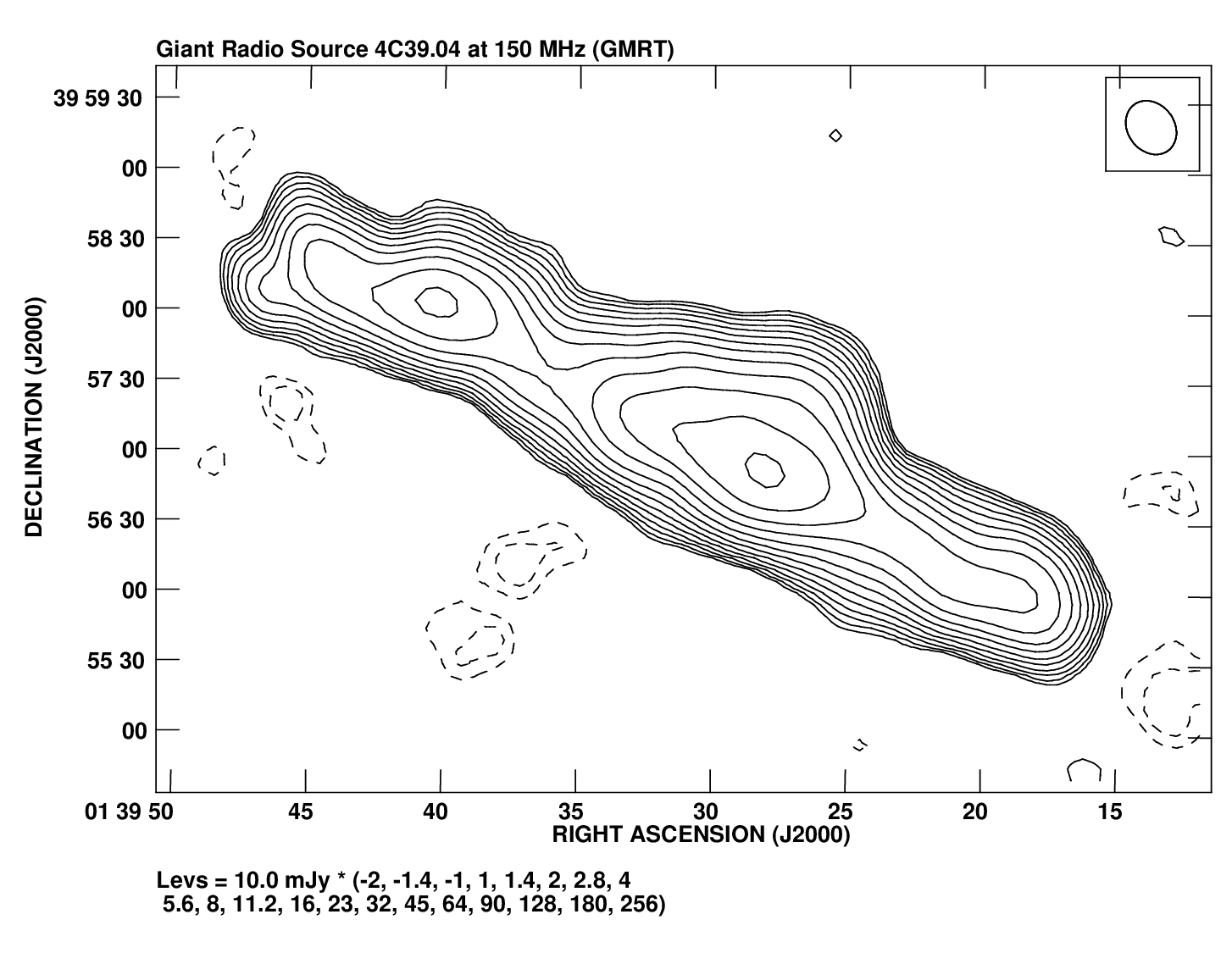}}
\bigskip
\centerline{\includegraphics[height=7cm]{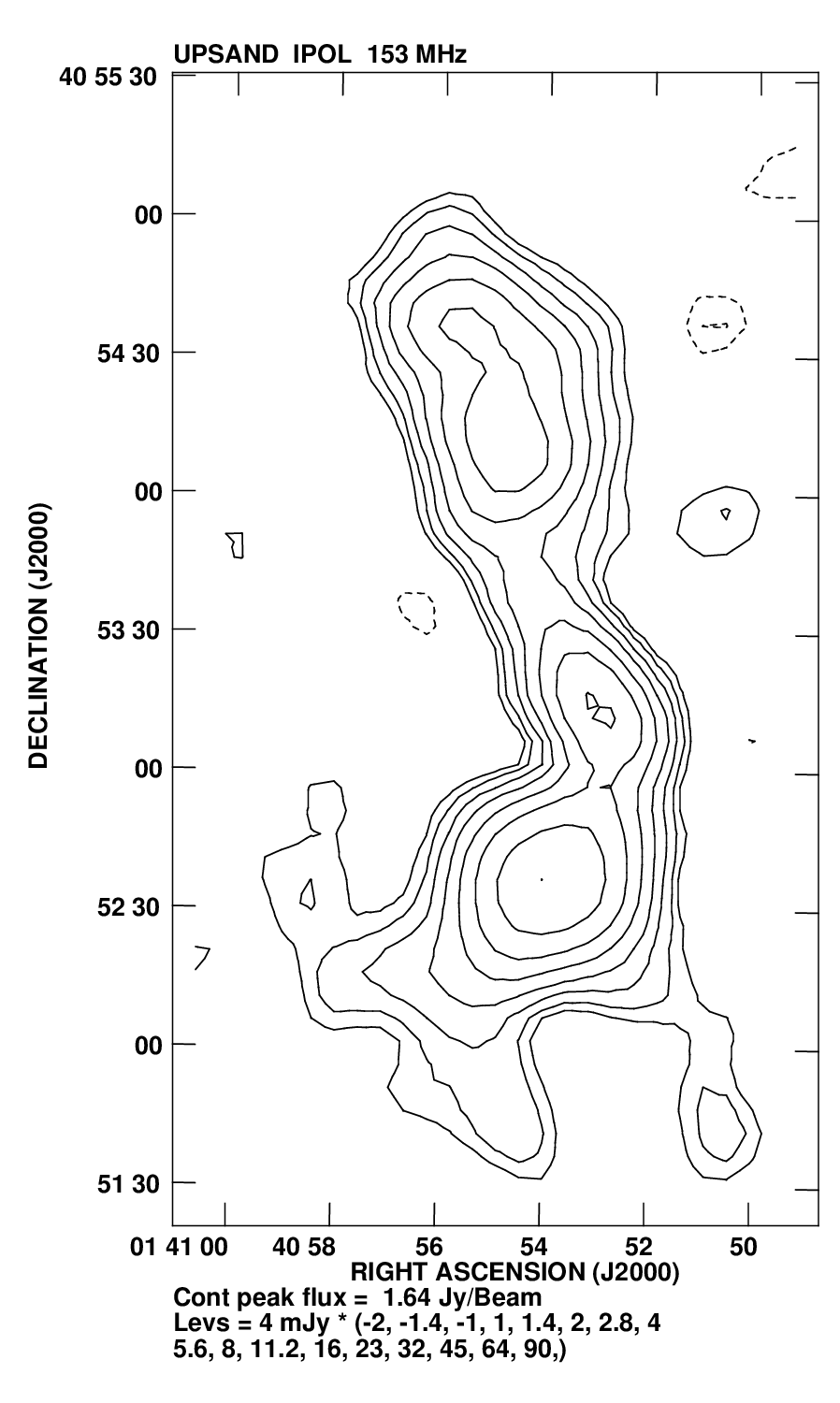}
     \qquad \includegraphics[height=7cm]{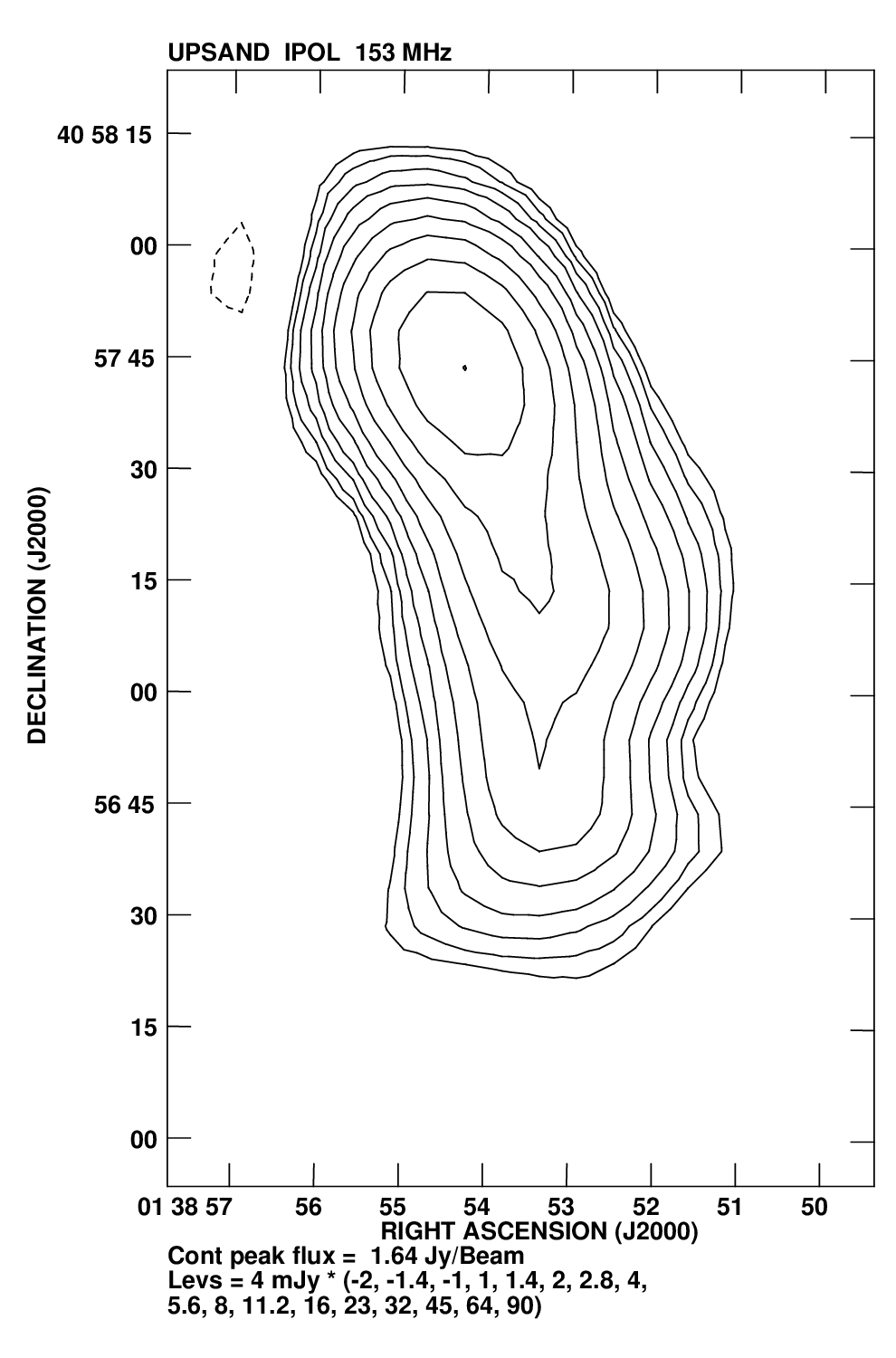}}
\caption{Three of the background sources discovered from the 150~MHz map of the
field of Ups And. {\it Top:} Giant radio source 4C39.04. This image at 150 MHz
is the highest sensitivity and also highest resolution image at this frequency.
The angular size is about 7 arcmin. {\it Lower left:} GMRT J0140+4053, from the
morphology, this resembles a low luminous radio galaxy with diffuse lobe on
either side. The central component does not look compact. {\it Lower right:}
GMRT J0138+4057, this is a candidate for a head--tail source.}
\label{extended_upsand}
\end{figure}

\section{Results and Discussion}

Apart from the hundreds of background sources which are largely unresolved, we
identified 4 extended objects of interesting morphology to follow-up at higher
frequencies. Below we discuss their characteristics from 150~MHz up to 5~GHz.

\textbf{GMRT J0137+4121}: This radio source is near the centre of the field and
has diffuse lobes on either side of a bright central component (see Fig.~1).
The central component has a peak flux of $\sim$ 80 mJy beam$^{-1}$ and an
integrated flux density of $\sim$ 120~mJy, which appears to be partially
extended and could be compact FRII radio source. There is a galaxy like object
seen in the Digitized Sky Survey some $7''$ away from the core. Investigation
of radio sources with similar morphology in the past has revealed a compact
double at the centre with diffuse lobes on either side. This indicates multiple
epochs of AGN activity where the diffuse outer lobes belong to the previous
epoch of AGN activity and the inner compact double corresponds to the most
recent epoch of activity (e.g.\ Jamrozy et al.\ 2007). Further observations
with GMRT at 610 MHz indeed revealed the possibility of a compact double at the
centre (see Fig.~1). However, when the source was observed with VLA at higher
radio frequencies with higher resolution, the central compact double morphology
was resolved to a core and one sided detached jet.  Further observations,
including determining the redshift of this object are necessary.

\textbf{GMRT J0138+4057}: The morphology of this source is typical of head-tail
radio sources (Lal \& Rao 2004), that are usually found in dense environments
(see Fig.~\ref{extended_upsand}). At 150~MHz the central component of this
source has a peak flux of $\approx 90$~mJy beam$^{-1}$.

\textbf{GMRT J0140+4053}: This source is very similar to GMRT J0137+4121,
though it is much more diffuse lacking a strong central component (see
Fig.~\ref{extended_upsand}). From the morphology, this resembles a low
luminous radio galaxy with diffuse lobe on either side. An optical galaxy
appears to be located at the position of the core. At 610~MHz we are able to
partially resolve a compact core with a peak flux of 13~mJy beam$^{-1}$.

\textbf{GMRT J0139+3956 (4C39.04)}: This is the well know Giant Radio Source
(GRS) 4C39.04. The 150~MHz map (Fig.~\ref{extended_upsand}) is the highest
resolution and highest sensitivity map at this frequency. The angular size of
this source is about 7~arcmin. The central component has a peak flux of
$\approx 680$~mJy beam$^{-1}$ at 150 MHz. It is important to study the spectral
index variation across the source at low radio frequencies (i.e.\ $<1$ GHz),
which will preferentially probe old populations of electrons. Archive data at
the GMRT exists for this source at L-band and at VLA up to 5 GHz. Our 610~MHz
observations with GMRT revealed a bright central core $\approx 32$~mJy
beam$^{-1}$. The large-scale structure of this source, showing relaxed lobes
has been reported by a number of authors (e.g.\ Hine 1979; Vigotti et al.\
1989). Our results enable us to further interpret the spectral shape of
this object. The core spectral index between 150~MHz and 4.8~GHz is 0.8, which
suggests the existence of a Steep Spectrum Core (SSC).

\section{Concluding Remarks}

Multifrequency GMRT and VLA observations have been presented and have helped
clarified the nature of 4C39.04 as a SSC. For these sources (apart from
4C39.04), high resolution optical data is required to determine the source
redshift and galaxy type, thus allowing for a global picture of the source. In
the case of GMRT J0138+4057, GMRT J0140+4053 higher resolution radio data is
required to resolve the possible core and like with GMRT J0137+4121 (using the
VLA at 4~GHz) allow the radio source nature to be determined.  In summary we
have been able to determine a detailed picture of a number of extended sources
detected at 150~MHz, quantifying their nature up to 4~GHz. This is just a small
number of extended sources from single pointing, but should be typically found
in the low frequency sky with a many of them of previously undetected. The GMRT
archive should be exploited before and used in conjunction with future low
frequency radio arrays.

\acknowledgements

We thank the staff of the GMRT who have made these observations possible. The
GMRT is run by the National Centre for Radio Astrophysics of the Tata Institute
of Fundamental Research.

\end{document}